*D. M. Golubchikov*
*K. E. Rumiantsev, Dr. of Eng. Sc., Prof*
*[Taganrog Institute of Technology - Southern Federal University]*
*Russia.*

# THE PROBABILISTIC MODEL OF KEYS GENERATION OF QKD SYSTEMS

***The probabilistic model of keys generation of QKD systems is proposed. The model includes all phases of keys generation starting from photons generation to states detection taking characteristics of fiber-optics components into account. The paper describes the tree of events of QKD systems. Equations are found for estimation of the effectiveness of the process of sifted keys generation as well as for bit-error probability and for the rate of private keys generation.***

Nowadays, the research of quantum cryptography transforms from theoretical models to practical implementations. The quantum key distribution is a branch of quantum cryptography in which scientists reach the most of practical results [1, 2]. Id Quantque (Switzerland)[3], MagiQ Technologies (USA)[4], SmartQuantum (France)[5], Quintessence Labs (Australia)[6] commercial companies offer first on the market quantum key distribution systems.

One of the main problems of quantum key distribution is low rate of private keys generation [7]. For example, Id 3100 Clavis2 system generates raw keys with the rate of about 500 bits per second when the length of quantum channel is shorter than 25 km [8]. Rates of keys generation of QKD systems of other companies are also not too high. It should be noted, that with increase of quantum channel length the rate of keys generation decreases.

Significant difference between the rates of raw and private keys generation is an interesting fact [9]. The rate of raw keys generation depends from the rate of photon generation, coefficient of photon transmission and photon detection efficiency. The rate of private keys generation depends from the rate of raw keys generation and bit-error probability. So the main stages of keys generation process are photon generation, photon transmission, basis selection and photon detection.

The tree of events of QKD systems is constructed taking into account the main stages of keys generation process (Figure 1).

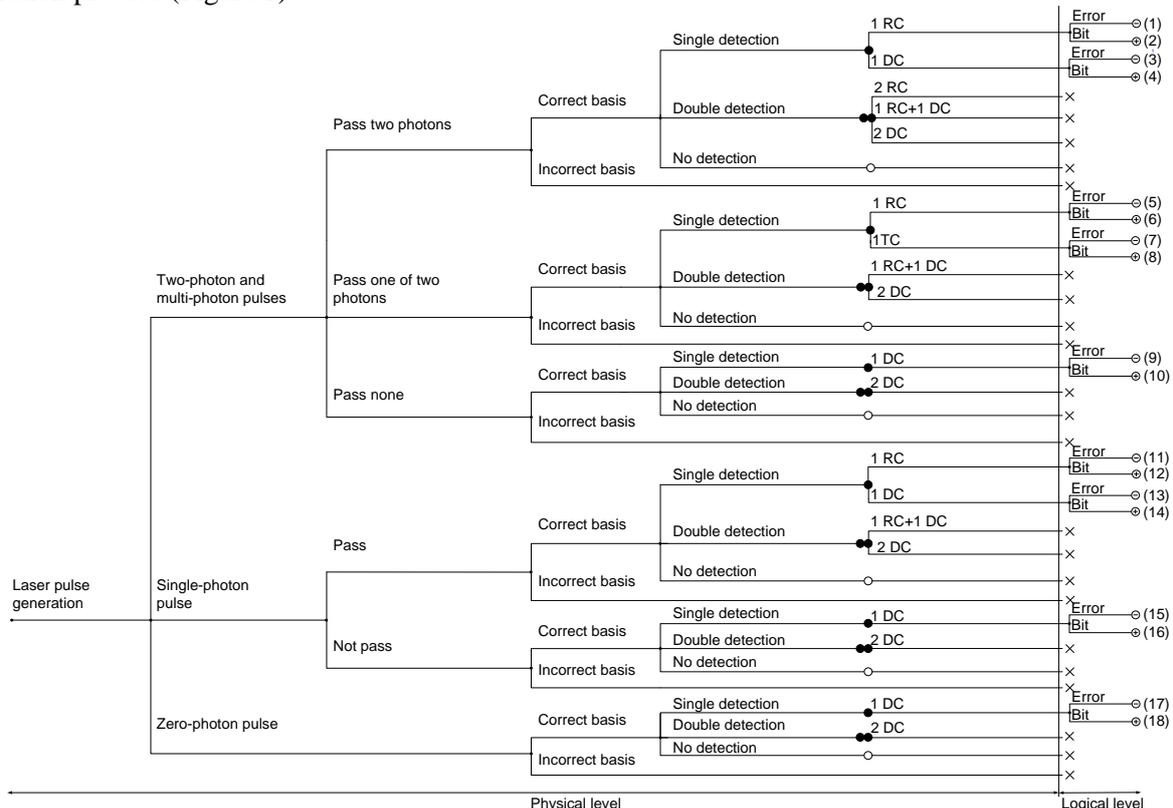

Figure 1. Tree of events of QKD systems.

The tree of events divides into physical and logical levels. The physical level consists of photon generation, photon transmission, basis selection and photon detection processes. The logical level consists of process of getting information about value of sifted keys bits and bit-error probability estimation process.

The root of tree is the event of laser pulse generation which has the probability $p_{si} = 1$. Each branch of tree corresponds to event that occurred at each stage of key generation. First junction of tree corresponds to the process of photon generation. Commercial QKD systems [3-5] use attenuated laser pulse for single photon generation. There are three branches because there are probability of zero-photon pulses $p_0(\mu)$, single-photon pulses probability $p_1(\mu)$ and two- and multi-photon pulses probability $p_2(\mu)$. $p_2(\mu)$ is combined of probabilities of pulses with two and more photons and is calculated as $p_2(\mu) = 1 - p_0(\mu) - p_1(\mu)$. For more precise calculation of the probability of multi-photon pulses expression $p_n(\mu) = \frac{\mu^n}{n!} \cdot \exp(-\mu)$ is used, where $\mu$ is average number of photons per pulse, $n$ is a quantity of photons per pulse. The average number of photons per pulse is calculated as $\mu(\alpha_{qc}, \alpha_{voaA}, \alpha_{voaB}) = \frac{E_{lp} \cdot f(\alpha_{qc}, \alpha_{voaA}, \alpha_{voaB})}{E_p}$, where $E_{lp}$ is energy of laser pulse, $E_p$ is energy of single photon, $\alpha_{qc}$ is a transfer ratio of quantum channel, $\alpha_{voaA}$ is a transfer ratio of Alice's variable optical attenuator, $\alpha_{voaB}$ is a transfer ratio of Bob's variable optical attenuator.

The average number of photons $\mu \in (0.1, 0.5)$ is used as an optimal level in QKD systems [10]. If $\mu > 0.5$ then the probability of multi-photon pulses increases and intercept security against photon number split attack decreases. The probability $p_2(0.5) \approx 0.09$ includes the probability of two-photon pulses of 0.076, so the probability of multi-photon pulses is lesser than 0.02. Figure 1 demonstrates the tree with 3 branches of events of photon generation process.

If alternative principle of photon generation process is used in given QKD system then other function of probability distribution will be used [11, 12].

Next junction of the tree at the path of non-zero photon pulse describes the photon transmission process. For single-photon pulses there are two possible branches: pass / not pass. For two-photon pulses there are three possible branches: pass two photons / pass one of two photons / pass none. For multi-photon pulses the number of possible branches is $n+1$. The probability of each one of them is a function of number of photons per pulse $p_m(n) = C_n^m \cdot (1 - p_{qc}(\alpha_{qc}))^{n-m} \cdot p_{qc}(\alpha_{qc})^m$, where $\alpha_{qc}$ is transfer ratio of quantum channel and $p_{qc}(\alpha_{qc})$ is probability for the photon to pass through the quantum channel, $m$ is quantity of transmitted photons, $n$ is the quantity of photons per pulse.

The basis selection process is defined by quantum protocol implementation. Commercial QKD systems [3-5] are using BB84 and SARG04 protocols. The probability of correct basis selection is $p_{cb} = p\{B|A\} = p\{0|0\} + p\{\pi/2|\pi/2\} + p\{0|\pi\} + p\{\pi/2|3\pi/2\}$. Therefore, the probability of incorrect basis selection is $1 - p_{cb}$. If all $p\{A\}$ where $A \in \{0, \pi/2, \pi, 3\pi/2\}$ are equal and all $p\{B\}$ where $B \in \{0, \pi/2\}$ are equal then $p_{cb} = 0.5$. The process of basis selection is positioned at physical level because it is implemented by fiber-optic phase modulator.

Next to last junction of the tree on physical level there is the photon detection process. It is well known that commercial QKD systems have two single photon detectors based on avalanche photodiodes. So branches refer to following events: no detection / single detection / double detection. These events describe how many detectors trigger when photon pulse arrives. Events are depicted on Figure 1 with ○ for no detection events, ● for single detection events and ●● for double detection events. Next junctions of the tree are related to the cause of detection. The single photon detectors based on avalanche photodiode have two causes of triggering: a dark count or a real count.

The real count probability is $p_i(\eta_j) = 1 - \exp(-p_a \cdot \eta_j \cdot i)$ where $i$ is the quantity of photons at entrance of single photon detector, $j$ is the ordinal number of the detector and $p_a$ is the probability of triggering the avalanche event that exceeds the threshold by each carrier. The dark count probability is

$p(dc_j) = 1 - \exp(-p_a \cdot N_d)$ where $N_d$ is the average number of dark carriers in the multiplication region of the $j$ detector [13].

Only single photon detection events are processed at logical level. All other events do not contain the information about the bit value of the keys.

As the result the two branches are available at the last stage: there is a correct bit or there is an error bit. Symbol $\oplus$ denotes events which lead to the correct bit of the shifted key as the result, and symbol $\ominus$ denotes events which have the error bit of the shifted key as the result. All events at the last stage are numbered.

In accordance with the tree of events of QKD systems all outcomes are united into two groups which both consist of two sub-groups:
1. Generating of the error bit
    a.  as the result of the dark count on single photon detector (events 3, 7, 9, 13, 15, 17),
    b.  as the result of the real count on single photon detector (events 1, 5, 11);
2. Generating of the correct bit
    a.  as the result of the dark count on single photon detector (events 4, 8, 10, 14, 16, 18),
    b.  as the result of the real count on single photon detector (events 2, 6, 12).

A bit of sifted key generation is the result of processing of each event.

The expression to calculation of the probability of event numbered k is
$$p_k = p_n(\mu) \cdot p_m(n) \cdot p_{cb} \cdot f(p(dc_j), p_i(\eta_j)) \cdot f(p_{err}, p_{bit}),$$
where $f(p(dc_j), p_i(\eta_j))$ is a function that describes the cause of detection and $f(p_{err}, p_{bit})$ is a function that describes a correctness of received bit.

The effectiveness of the process of sifted keys generation is
$$p_\Sigma(\mu, \alpha_{qc}, dc_1, dc_2, \eta_1, \eta_2) = p_{\Sigma err.dc}(\mu, \alpha_{qc}, dc_1, dc_2, \eta_1, \eta_2) + p_{\Sigma bit.dc}(\mu, \alpha_{qc}, dc_1, dc_2, \eta_1, \eta_2) + \\ + p_{\Sigma bit.rc}(\mu, \alpha_{qc}, dc_1, dc_2, \eta_1, \eta_2) + p_{\Sigma err.rc}(\mu, \alpha_{qc}, dc_1, dc_2, \eta_1, \eta_2),$$
where $p_{\Sigma err.dc}$ is sum of probabilities of 1a sub-group events, $p_{\Sigma bit.dc}$ is sum of probabilities of 2a sub-group events, $p_{\Sigma err.rc}$ is sum of probabilities of 1b sub-group events and $p_{\Sigma bit.rc}$ is sum of probabilities of 2b sub-group events.

The effectiveness of the process of sifted keys generation describes QKD system capability to form a bit of sifted key from laser pulse.

The value of bit-error probability is needed to estimate the rate of private keys generation.

The bit-error probability is
$$p_{err} = \frac{p_{\Sigma err.dc}(\mu, \alpha_{qc}, dc_1, dc_2, \eta_1, \eta_2) + p_{\Sigma err.rc}(\mu, \alpha_{qc}, dc_1, dc_2, \eta_1, \eta_2)}{p_\Sigma(\mu, \alpha_{qc}, dc_1, dc_2, \eta_1, \eta_2)}.$$

So the rate of private keys generation is
$$v_\Sigma(\mu, \alpha_{qc}, dc_1, dc_2, \eta_1, \eta_2) = v \cdot p_\Sigma(\mu, \alpha_{qc}, dc_1, dc_2, \eta_1, \eta_2) \cdot \varepsilon(p_{err}),$$
where $v$ is effective frequency of laser pulse generation and $\varepsilon(p_{err})$ is a coefficient of private keys length shortening after error correction and privacy amplification [14].

## Conclusion

All commercial QKD system structures are analysed in order to construct the tree of events of QKD systems. The tree of events represents the main stages of keys generation process in QKD systems like process of photon generation, photon transmission, basis selection and photon detection. Probabilistic model based on the tree of events is used for estimation of influence of the characteristics of fiber-optic components to QKD system characteristics. The model may be used at the development stage of new QKD system for estimation of system-limiting characteristics based on known characteristics of fiber-optic components. The model also may be used to make requirements to the fiber-optic components for customized QKD systems.

## Acknowledgment

A part of this research is supported by Federal Program "Research and scientific-pedagogical personnel of innovation in Russia" for 2009 – 2013 years (No.NK-289p-4).